\documentclass[twocolumn,amsmath,amssymb]{revtex4}

\usepackage{graphicx}
\usepackage{dcolumn}
\usepackage{bm}

\begin{document}


\title{Newtonian gravitational deflection of light revisited\\~\\
(http://xxx.if.usp.br/abs/physics/0508030) }


\author{Domingos S.L. Soares}
 \email{dsoares@fisica.ufmg.br}
 \homepage{http://www.fisica.ufmg.br/~dsoares}
\affiliation{Physics Department, ICEX, UFMG, C.P. 702,\\
30.123-970, Belo Horizonte, MG
Brazil
}

\date{\today}

\begin{abstract}
The angle of deflection of a light ray by the gravitational field of the Sun, 
at grazing incidence, is calculated by strict and straightforward classical 
Newtonian means using the corpuscular model of light. The calculation is presented 
in the historical and scientific contexts of Newton's {\it Opticks} and of modern 
views of the problem.
\end{abstract}

\maketitle

\section{Introduction}

Isaac Newton's {\it Opticks} \cite{IN1} --- whose last edition, the fourth, was 
published  in 1730 --- is entirely dedicated to the description of experiments 
in optics. The subtitle reads {\it``A Treatise of the Reflections, 
Refractions, Inflections \& Colours of Light''}. It is composed by three 
Books. Book One synthesizes the previous knowledge on the properties of 
light and is presented as a series of definitions, axioms and propositions.
Books Two and Three are devoted to Newton's experiments and to the many 
discoveries he made therefrom. Book Three ends with a series of 31 queries. 
In Newton's words: {\it ``When I made the foregoing Observations, I design'd 
to repeat most of them with more care and exactness, (...) But I was then 
interrupted, and cannot now think of taking these things into farther 
Consideration. (...) I shall conclude with proposing only some Queries, 
in order to a farther search to be made by others.''} \cite{IN2}.

The modern concept of the gravitational deflection of light is very much an 
Einsteinian idea \cite{overb,pais}, though 
traces of the idea can be found in Newton's Query 1, namely, {\it 
``Do not Bodies act upon Light at a distance, and by their action bend its 
Rays, and is not this action (c\ae teris paribus) strongest at the least 
distance?''}. In conjunction with its corpuscular model of light, which is 
explicitly stated in Query 29 as {\it ``Are not the Rays of Light very small 
Bodies emitted from shining Substances?''}, one has the whole scenario for 
a Newtonian calculation of the gravitational deflection of light by a massive 
body. 

However there is not such a discussion in the {\it ``Opticks''}. One should 
remember that Newton was entirely concerned with the properties of light 
behavior in his daily laboratory experiments.

In the next section I shall pursue on such a scenario and calculate the 
Newtonian deflection of a light ray at grazing incidence in the solar limb.
The result is precisely the same as obtained with modern space-time curvature   
calculations except for a factor of two. For example, Weinberg \cite{weinb} obtains 
the general solution for grazing incidence at the Sun. The solution holds for 
various theories of gravity including General Relativity and Newton's gravity. 
Denoting the angles of deflection for these cases by $\delta_{GR}$ 
and $\delta_{NG}$, Weinberg's general expression yields

\begin{equation}
\label{eq:weinbGR}
\delta_{GR} = \frac{4GM_S}{c^2R_S} = 1.75~{\rm arcsec}
\end{equation}

and

\begin{equation}
\label{eq:weinbN}
\delta_{NG} = \frac{2GM_S}{c^2R_S} = \frac{1}{2}\delta_{GR} ,
\end{equation}

where $G$, $c$, $M_S$ and $R_S$ are the gravitational constant, the speed of 
light, the Sun's mass and the Sun's radius, respectively. 

Einstein's first calculation of the gravitational deflection of light, in 
1911 (see \cite{overb}, for a historical account and scientific 
references), was performed using the Equivalence Principle and the equivalent 
mass-energy of a photon. The calculation yielded $\delta_{NG}$. Only in his 
second calculation, published in 1916, where he included the effect of 
space-time curvature, he obtained a value twice as large as his first 
calculation, i.e., $\delta_{GR}$ \cite{overb}.

Before going on with the classical derivation of $\delta_{NG}$, it is 
worthwhile to mention an interesting note by I. Bernard Cohen, in the Preface 
of the modern edition of {\it Opticks} used here. On page xxxiii, he 
comments about the Queries: {\it ``To be sure, the 
speculations of the Opticks were not hypotheses, at least to the extent that 
they were framed in questions. Yet if we use Newton's own definition, that 
`whatever is not deduced from the phenomena is to be called an hypothesis' 
they are hypotheses indeed. The question form may have been adopted in order 
to allay criticism, but it does not hide the extent of Newton's belief. For 
every one of the Queries is phrased in the negative! Thus Newton does not ask 
in truly interrogatory way (Qu. 1) : `Do Bodies act upon Light at a 
distance... ?' --- as if he did not know the answer. Rather, he puts it: `Do 
not Bodies act upon Light at a distance... ?' --- as if he knew the answer 
well --- `Why, of course they do!' ''}

\section{Deflection of a light ray by the Sun}

Taking Queries 1 and 29 above at their face values, one may proceed 
to calculate the deflection of a light ray by the gravitational field of a 
massive body. 

\begin{figure}
\includegraphics[width=8cm]{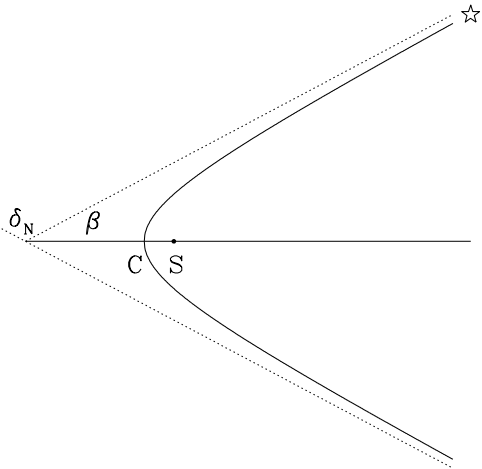}
\caption{\label{fig:hypb}The light ray from a star follows an unbound 
hyperbolic orbit about the Sun. For deflection on grazing incidence, the 
distance CS is R$_S$, the solar radius. For illustration purposes, the 
bending is greatly exaggerated.}
\end{figure}

A light ray, from a distant star, under the Sun's gravitational force field 
describes the usual central force hyperbolic orbit. The deflection of the 
light ray is illustrated in Fig.~\ref{fig:hypb}, with the bending greatly 
exaggerated for a better view of the angle of deflection.

At grazing incidence through the solar limb, the distance CS is the solar 
radius $R_S$. The angle $\beta$ of the asymptote to the hyperbole of 
eccentricity $\epsilon$ is given by \cite{sym}

\begin{equation}
\label{eq:beta}
\cos\beta = \frac{1}{\epsilon}.
\end{equation}

The angle of deflection of the light ray, $\delta_N$, is shown in 
the figure and is 

\begin{equation}
\label{eq:del}
\delta_N = \pi - 2\beta.
\end{equation}

For a light particle with mass $m$, total energy $E$ and angular momentum 
$L$, with respect to the Sun's center S, the eccentricity is 

\begin{equation}
\label{eq:eps}
\epsilon = \left( 1+\frac{2EL^2}{G^2m^3M_S^2}\right)^{1/2}.
\end{equation}

The constants of motion $E$ and $L$ are easily evaluated at point C. 
Assuming that the particle speed at that point is $v$ one has 

\begin{equation}
E = \frac{1}{2}mv^2 - \frac{GmM_S}{R_S}
\end{equation}

and 

\begin{equation}
L = mvR_S.
\end{equation}

Since the light particle speed is very large --- much larger than the 
escape speed at the solar radius --- one can neglect the change 
in its magnitude and assume that the gravitational field changes only 
the velocity direction. Making $v\equiv c$, the light speed in vacuum, the 
eccentricity is written as

\begin{equation}
\label{eq:eps1}
\epsilon = \left[ 1+\frac{\left(c^2-2GM_S/R_S\right)c^2R_S^2}{G^2M_S^2}\right]^{1/2}.
\end{equation}

As expected, the light particle mass $m$ cancels out above. Also, 
the second term inside the parentheses is readily 
verified to be much smaller than $c^2$. Thus, the  eccentricity may be 
simplified to

\begin{equation}
\label{eq:eps2}
\epsilon = \frac{c^2R_S}{GM_S} = 4.7\times10^5 \gg 1.
\end{equation}

The angle of deflection, given by eq. \ref{eq:del}, can be written, with 
the aid of eqs. \ref{eq:beta} and \ref{eq:eps2} as 

\begin{equation}
\label{eq:del1}
\delta_N = \pi - 2\cos^{-1}\left(\frac{GM_S}{c^2R_S}\right)
\end{equation}

and since $x\equiv GM_S/c^2R_S \ll 1$, one can expand $\cos^{-1}x$ 
in a Taylor's series:

\begin{equation}
\label{eq:ser}
\cos^{-1}x=\frac{\pi}{2} - \sin^{-1}x=\frac{\pi}{2} - \left(x + 
\frac{1}{2}\frac{x^3}{3}+...\right).
\end{equation}

Considering only the first term inside the parentheses, 
$\delta_N$ in eq. \ref{eq:del1} takes the form of Weinberg's 
Newtonian deflection given by eq. \ref{eq:weinbN}:

\begin{equation}
\label{eq:delN}
\delta_{N} = \frac{2GM_S}{c^2R_S}.
\end{equation}

In conclusion, there is still the question of why Newton did not 
discuss the possibility of light ray deflection by a massive heavenly 
body. Of course, he was well acquainted with the relevant astronomical 
observations. Solar eclipses were certainly of his knowledge and could 
certainly motivate digressions on the gravitational bending of light. 
Nevertheless, there is nothing about such issues in the {\it Opticks}. 
As mentioned above, his main concerns were with the laboratory and 
daily-life behavior of light. Incidentally, North \cite{nor} states that the {\it 
``likelihood of such an effect had previously been maintained by Newton and 
Laplace...''} but gives no references. To my knowledge, the first
Newtonian approach to the problem of the gravitational bending of light was
undertaken by Johann G. von Soldner, in early XIX century. Soldner's pioneering
work is brilliantly reviewed by Stanley L. Jaki \cite{jaki},
who provided also an English translation of his article,
submitted for publication in 1801 and printed in 1804. 

\begin{acknowledgments}
I wish to thank Drs. Carlos Heitor D'\'Avila Fonseca, Ant\^onio Claret dos Santos
and Jos\'e Francisco de Sampaio for helpful comments.
After finishing the manuscript, my attention was drawn to a section of an 
electronic book entitled {\it ``Reflections on Relativity''}, which contains 
a more detailed account of the subject treated here. The Reader is referred to that 
book section at the URL {\tt http://www.mathpages.com/rr/s6-03/6-03.htm}.

\end{acknowledgments}


\begin{thebibliography}{}
\bibitem{jaki} S.L. Jaki, {\it Johann Georg von Soldner and the Gravitational
Bending of Light, with an English Translation of His Essay on It Published
in 1801}, 1978, Foundations of Physics, {\bf 8}, pp. 927-950
\bibitem{IN1} I. Newton, {\it Opticks}, (Dover, New York, 1979).
\bibitem{IN2} Ibid., pp. 338-339.
\bibitem{overb} D. Overbye, {\it Einstein in Love: A Scientific Romance}, 
(Viking Press, New York, 2000), chapters 14 and 25.
\bibitem{pais} A. Pais, {\it Subtle Is the Lord: The Science and the Life of 
Albert Einstein}, (Oxford University Press, Oxford, 1983), chapter 11.
\bibitem{weinb} S. Weinberg, {\it Gravitation and Cosmology}, (Wiley, New 
York, 1972), p. 188.
\bibitem{sym} K. Symon, {\it Mechanics}, (Addison-Wesley, Reading, 1967), 
p. 130.
\bibitem{nor} J.D. North, {\it The Measure of the Universe: A History of Modern 
Cosmology}, (Dover, New York 1990), p. 68.
\end{thebibliography}
\end{document}